\documentclass[usegraphicx,usenatbib,useAMS,twocolumn]{mn2e}

\usepackage{epsfig}
\usepackage{graphicx}
\usepackage{amsmath,bm}
\usepackage{color}
\usepackage{amssymb}
\usepackage{times}

\def\jnl@style{\it}
\def\aaref@jnl#1{{\jnl@style#1}}

\def\aaref@jnl#1{{\jnl@style#1}}

\def\aj{\aaref@jnl{AJ}}                   
\def\araa{\aaref@jnl{ARA\&A}}             
\def\apj{\aaref@jnl{ApJ}}                 
\def\apjl{\aaref@jnl{ApJ}}                
\def\apjs{\aaref@jnl{ApJS}}               
\def\ao{\aaref@jnl{Appl.~Opt.}}           
\def\apss{\aaref@jnl{Ap\&SS}}             
\def\aap{\aaref@jnl{A\&A}}                
\def\aapr{\aaref@jnl{A\&A~Rev.}}          
\def\aaps{\aaref@jnl{A\&AS}}              
\def\azh{\aaref@jnl{AZh}}                 
\def\baas{\aaref@jnl{BAAS}}               
\def\jrasc{\aaref@jnl{JRASC}}             
\def\memras{\aaref@jnl{MmRAS}}            
\def\mnras{\aaref@jnl{MNRAS}}             
\def\pra{\aaref@jnl{Phys.~Rev.~A}}        
\def\prb{\aaref@jnl{Phys.~Rev.~B}}        
\def\prc{\aaref@jnl{Phys.~Rev.~C}}        
\def\prd{\aaref@jnl{Phys.~Rev.~D}}        
\def\pre{\aaref@jnl{Phys.~Rev.~E}}        
\def\prl{\aaref@jnl{Phys.~Rev.~Lett.}}    
\def\pasp{\aaref@jnl{PASP}}               
\def\pasj{\aaref@jnl{PASJ}}               
\def\qjras{\aaref@jnl{QJRAS}}             
\def\skytel{\aaref@jnl{S\&T}}             
\def\solphys{\aaref@jnl{Sol.~Phys.}}      
\def\sovast{\aaref@jnl{Soviet~Ast.}}      
\def\ssr{\aaref@jnl{Space~Sci.~Rev.}}     
\def\zap{\aaref@jnl{ZAp}}                 
\def\nat{\aaref@jnl{Nature}}              
\def\iaucirc{\aaref@jnl{IAU~Circ.}}       
\def\aplett{\aaref@jnl{Astrophys.~Lett.}} 
\def\apspr{\aaref@jnl{Astrophys.~Space~Phys.~Res.}}
\def\bain{\aaref@jnl{Bull.~Astron.~Inst.~Netherlands}} 
\def\fcp{\aaref@jnl{Fund.~Cosmic~Phys.}}  
\def\gca{\aaref@jnl{Geochim.~Cosmochim.~Acta}}   
\def\grl{\aaref@jnl{Geophys.~Res.~Lett.}} 
\def\jcp{\aaref@jnl{J.~Chem.~Phys.}}      
\def\jgr{\aaref@jnl{J.~Geophys.~Res.}}    
\def\jqsrt{\aaref@jnl{J.~Quant.~Spec.~Radiat.~Transf.}}
\def\memsai{\aaref@jnl{Mem.~Soc.~Astron.~Italiana}}
\def\nphysa{\aaref@jnl{Nucl.~Phys.~A}}   
\def\physrep{\aaref@jnl{Phys.~Rep.}}   
\def\physscr{\aaref@jnl{Phys.~Scr}}   
\def\planss{\aaref@jnl{Planet.~Space~Sci.}}   
\def\procspie{\aaref@jnl{Proc.~SPIE}}   

\title[Measuring line-of-sight-dependent Fourier-space clustering using FFTs]{Measuring line-of-sight-dependent Fourier-space clustering using FFTs}

\author[D. Bianchi et al.]{\parbox{\textwidth}{ 
Davide Bianchi\thanks{davide.bianchi@port.ac.uk}, 
H\'ector Gil-Mar\'in,
Rossana Ruggeri \&
Will J. Percival
}
\vspace*{4pt} \\
Institute of Cosmology \& Gravitation, University of Portsmouth, Dennis Sciama Building, Portsmouth PO1 3FX, UK
}

\def\gs{\mathrel{\raise1.16pt\hbox{$>$}\kern-7.0pt
\lower3.06pt\hbox{{$\scriptstyle \sim$}}}}         
\def\ls{\mathrel{\raise1.16pt\hbox{$<$}\kern-7.0pt 
\lower3.06pt\hbox{{$\scriptstyle \sim$}}}}         

\begin{document}
\maketitle

\begin{abstract} 

Observed galaxy clustering exhibits local transverse statistical isotropy around the line-of-sight (LOS). The variation of the LOS across a galaxy survey complicates the measurement of the observed clustering as a function of the angle to the LOS, as fast Fourier transforms (FFTs) based on Cartesian grids, cannot individually allow for this. Recent advances in methodology for calculating LOS-dependent clustering in Fourier space include the realization that power spectrum LOS-dependent moments can be constructed from sums over galaxies, based on approximating the LOS to each pair of galaxies by the LOS to one of them. We show that we can implement this method using multiple FFTs, each measuring the LOS-weighted clustering along different axes. The $N\log N$ nature of FFTs means that the computational speed-up is a factor of $>1000$ compared with summing over galaxies. This development should be beneficial for future projects such as DESI and {\it Euclid} which will provide an order of magnitude more galaxies than current surveys.  

\end{abstract}

\begin{keywords}
cosmology: theory - large-scale structure of Universe 
\end{keywords}

\section{Introduction}\label{sec:intro}
Although the Universe is predicted to be statistically homogeneous and isotropic, observational effects including the Alcock-Paczynsky effect \citep[AP;][]{alcock1979} and redshift-space distortions \citep[RSD;][]{kaiser1987} mean that the observed clustering, when translated into comoving coordinates using a fiducial distance-redshift relation exhibits local transverse statistical isotropy around the line-of-sight (LOS).  The key measurement to be made from a galaxy survey is consequently the clustering as a function of the angle to the LOS. If we consider the clustering in configuration-space, then the base `unit' is a pair of galaxies, and it is common to treat a pair as having a single LOS, usually defined as the direction to the pair centre.  Any effects because the galaxies within the pair have different LOSs are called `wide-angle effect' \citep{szalay1998, szapudi2004} and are small of the scales of interest \citep{beutler2012, samushia2012, yoo2015}.  Thus in configuration space, measuring clustering with respect to the LOS can be easily incorporated into pair-counting algorithms \citep{landy1993} with a different LOS for each pair.

In Fourier-space, dealing with the varying LOS is more difficult, as fast Fourier methods do not, in general, allow for the variation of LOS. One option is to use a basis built up from spherical harmonics and Bessel functions, which naturally separates clustering with respect to the varying LOS \citep{fisher1994c, heavens1995}. In recent works, \citet{yamamoto2006} and \citet{blake2011} considered the Fourier decomposition as a sum over pairs of galaxies, and showed that this can be simplified (and speeded up) by assuming that the LOS to the pair is equivalent to the LOS to a single galaxy (the method is described in \S~\ref{sec:review}). This approximately doubles the `wide-angle effect' \citep{samushia2015}, but that is small anyway. In this Letter we consider how to implement the transform with this approximation, showing that we can use multiple fast Fourier transforms (FFTs) to perform this sum for power-law moments in $\mu \equiv \hat{\bf k}\cdot\hat{\bf r}_{\rm LOS}$, the cosine of the angle to the LOS (this is described in \S~\ref{sec:algorithm}). In \S~\ref{sec:tests} we present the results of tests of three implementations of the method, summing over galaxies, grid cells or using FFTs. We show that they provide consistent results, and compare the computational burden of each. By decomposing any moment into a sum over Legendre polynomials, we can construct any power spectrum moment using this method (\S~\ref{sec:moments}). Such developments are necessary as one often wants to measure the power spectrum moments, not only in the data, but also in a large numbers of mock catalogues used to estimate and test for errors: for example, \citet{anderson2014} analysed the Baryon Oscillation Spectroscopic Survey (BOSS; \citealt{dawson2012}) data and 1000 mock catalogues. Thus the computational burden of measuring LOS-dependent clustering is high.

\section{Method}\label{sec:review}
We start by defining the function \citep{feldman1994}, 
\begin{equation}
F({\bf r})=\frac{w({\bf r})}{I^{1/2}} [n({\bf r}) - \alpha n_s({\bf r})],
\label{eq:FKP_factor}
\end{equation}
where $n$ and $n_s$ are, respectively, the observed number density of galaxies and the number density of a synthetic catalog of {\it randoms}, Poisson sampled with the same mask and selection function as the survey with no other cosmological correlations, and $w$  is the weight. $\alpha$ normalizes the weighted random catalogue to match the weighted galaxy catalogue. The factor $I$ normalizes the amplitude of the observed power in accordance with its definition in a universe with no survey selection, $I\equiv \int d{\bf r}\,w^2{\bar{n}}^2({\bf r})$.
From Eq.~(\ref{eq:FKP_factor}) we can define the multipole power spectrum estimator as \citep{feldman1994, yamamoto2006},
\begin{align}  
\nonumber\hat{P}_\ell({ k})=& \ \frac{(2\ell+1)}{I}\int \frac{d\Omega_k}{4\pi}\, \left[ \int d{\bf r}_1\,\int d{\bf r}_2\,F({\bf r}_1)F({\bf r}_2)\right.\\
\label{eq:P_estimator} &\times \left.e^{i{\bf k}\cdot({\bf r}_1-{\bf r}_2)}\mathcal{L}_\ell(\hat{\bf k}\cdot\hat{\bf r}_h)-P_\ell^{\rm noise}({\bf k})\right] \ ,
\end{align}
where ${\bf r}_h\equiv ({\bf r}_1+{\bf r}_2)/2$ denotes the LOS of the pair of galaxies ${\bf r}_1$  and ${\bf r}_2$, $d\Omega_k$ is the solid angle element in $k$-space, $\mathcal{L}_\ell$ is the $\ell-${th} order Legendre polynomial and $P_\ell^{\rm noise}$ is the shot noise term given by
\begin{equation}
P_\ell^{\rm noise}({\bf k}) = (1+\alpha)\int d{\bf r}\, \bar{ n} ({\bf r})w^2({\bf r})\mathcal{L}_\ell (\hat{\bf k}\cdot\hat{\bf r}) \ .
\end{equation}
For multipoles of order $\ell>0$, $P_\ell^{\rm noise}\ll \hat{P_\ell}$, and consequently the shot noise correction is negligible.

Denoting the number of $k$-modes that we want to evaluate by $N_k$ and the number of elements that we use to perform the integral over ${\bf r}_1$ or ${\bf r}_2$ by $N$, we see that the computation of Eq.~(\ref{eq:P_estimator}) will be of order $N_k\times N^2$, as the integrals in ${\bf r}_1$ and ${\bf r}_2$ are not separable. In effect this approach performs a pair-wise clustering analysis and translates into Fourier-space. As $N$ increases the total time needed to evaluate Eq.~(\ref{eq:P_estimator}) grows dramatically. 

The FKP-estimator \citep{feldman1994} uses the fact that the monopole is independent of the LOS, so the ${\bf r}_i$ integrals are separable and FFTs are trivial to apply. Consequently, the $N_k\times N^2$ process becomes a $N_k\log(N)$ one, which it is easier to handle: here $N$ is the number of grid cells at which we sample $F$, so for a FFT $N=N_k$. This estimator has been successfully applied in many galaxy surveys to estimate the power spectrum and bispectrum monopoles \citep[see e.g.][and references therein]{gil-marin2015}.

The Yamamoto estimator \citep{yamamoto2006, beutler2014} keeps the relevant LOS information by approximating the LOS of each pair of galaxies with the LOS of one of the two galaxies, $\mathcal{L}_\ell(\hat{\bf k}\cdot\hat{\bf r}_h)\simeq\mathcal{L}_\ell(\hat{\bf k}\cdot\hat{\bf r}_2)$, which yields
\begin{eqnarray}
\label{eq:P_yama}
\nonumber{\hat P_\ell^{\rm Yama} }(k)& =& \frac{(2\ell + 1)}{I} \int \frac{d\Omega_k}{4\pi}\, \left[ \int d{\bf r}_1\, F({\bf r}_1) e^{i{\bf k}\cdot{\bf r}_1}\right.\\
\nonumber &\times& \left.\int d{\bf r}_2\, F({\bf r}_2) e^{-i{\bf k}\cdot {\bf r}_2}\mathcal{L}_\ell({\hat {\bf k}}\cdot {\hat {\bf r}_2})-P_\ell^{\rm noise}({\bf k})\right].\\
\end{eqnarray}
This is a reliable approximation on the scale of interest, which clearly improves on assuming a single fixed LOS for the whole survey for $l>0$, but will eventually break down at very large scales \citep{samushia2015, yoo2015}.  The integrals over ${\bf r}_1$ and ${\bf r}_2$ in Eq.~(\ref{eq:P_yama}) are separable, so $\hat P_\ell^{\rm Yama}$ becomes a $N_k\times N$ process if the integrals are solved using sums (as in \citealt{beutler2014}).  In this Letter, we show that the efficiency of this estimator can be further improved by making use of FFT algorithms, such as \textsc{fftw}\footnote{Fastest Fourier Transform in the West: http://fftw.org}.

\section{FFT implementation}\label{sec:algorithm}

Here we show how to write the Yamamoto algorithm in terms of $N_k\log(N)$ processes for any multipoles.  For simplicity and with no loss of generality, we focus on the monopole (which, as discussed in \S \ref{sec:review}, reduces to the standard FKP description), the quadrupole and the hexadecapole. We proceed by defining the convenient function,
\begin{equation}
\label{eq:An}
A_n({\bf k}) = \int d{\bf r}\, {(\hat{\bf k}\cdot\hat{\bf r})}^n F({\bf r})e^{i{\bf k}\cdot{\bf r}}.
\end{equation}
With this,  Eq.~(\ref{eq:P_yama}) reads,
\begin{align}
\hat{P}_0^{\rm Yama}({ k}) =& \ \frac{1}{I} \int \frac{d\Omega_k}{4\pi}\, \left[ A_0({\bf k}) A^*_0({\bf k})\right] -P_0^{\rm noise}\\
\hat{P}_2^{\rm Yama}({ k}) =& \ \frac{5}{2I} \int \frac{d\Omega_k}{4\pi}\, A_0({\bf k}) \left[3A^{*}_2({\bf k})- A^*_0({\bf k}) \right],\\
\nonumber\hat{P}_4^{\rm Yama}({ k}) =& \ \frac{9}{8I}\int\frac{d\Omega_k}{4\pi} A_0({\bf k})\left[ 35A_4^*({\bf k})-30A_2^*({\bf k})\right.\\
&+\left.3A_0^*({\bf k}) \right] . 
\end{align}
Note that the expressions for $A_2$ and $A_4$ include a $k$-dependent term $(\hat{\bf k}\cdot\hat{\bf r})^n$ in the integrand, which means that in this form Fourier transforms cannot directly be applied. This is the standard problem of dealing with a varying LOS across a survey.  However, by means of the trivial decomposition 
\begin{equation}  \label{eq:kdotr}
\hat{\bf k}\cdot\hat{\bf r}=\frac{k_x r_x+k_y r_y+ k_z r_z}{kr} \ ,
\end{equation}
$A_2$ can be easily re-written into a combination of smaller building blocks,
\begin{align}
\label{eq:A2expansion}
\nonumber A_2({\bf k})=& \ \frac{1}{k^2}\left\{k_x^2B_{xx}({\bf k})+k_y^2B_{yy}({\bf k})+k_z^2B_{zz}({\bf k}) \right.\\
&+\left.2\left[  k_xk_yB_{xy}({\bf k})+ k_xk_zB_{xz}({\bf k})+ k_yk_zB_{yz}({\bf k})\right]\right\} \ ,
\end{align}
where
\begin{equation}
\label{Bij_def}B_{ij}({\bf k}) \equiv \int d{\bf r}\, \frac{r_i r_j}{r^2}F({\bf r})e^{i{\bf k}\cdot{\bf r}} \ .
\end{equation}
Similarly, for $A_4$ we obtain,
\begin{align}\label{eq:A4expansion}
\nonumber A_4({\bf k}) =& \ \frac{1}{k^4} \big\{k_x^4 C_{xxx}+k_y^4 C_{yyy}+k_z^4C_{zzz} \\
\nonumber &+ 4\left[ k_x^3 k_y C_{xxy} + k_x^3 k_z C_{xxz} + k_y^3 k_x C_{yyx} \right.\\
\nonumber &+ \left.k_y^3 k_z C_{yyz} + k_z^3 k_x C_{zzx} + k_z^3 k_y C_{zzy} \right] \\
\nonumber &+ 6\left[ k_x^2k_y^2 C_{xyy} + k_x^2 k_z^2 C_{xzz} + k_y^2 k_z^2 C_{yzz}\right] \\
&+ 12k_xk_yk_z\left[ k_x C_{xyz} + k_y C_{yxz} + k_z C_{zxy}\right]\big\} \ ,
\end{align}
where
\begin{equation}
\label{Cij_def}C_{ijl}({\bf k}) \equiv\int d{\bf r}\,\frac{r_i^2 r_j r_l}{r^4}F({\bf r})e^{i {\bf k}\cdot {\bf r}} \ .
\end{equation}
$A_0$, $B_{ij}$ and $C_{ijl}$ are all $N_k\log(N)$ processes by the use of any FFT algorithm. This provides the value of monopole, quadrupole and hexadecapole with only $1$, $7\, (=1+6)$ and $22\, (=1+6+15)$ FFTs, respectively. Similar decompositions are possible for higher order multipoles.

It is important to remark that, from an analytical point of view, the above decomposition is completely equivalent to Eq.~(\ref{eq:P_yama}), i.e. it does not involve any further approximation.
In essence, the symmetry encoded in the Yamamoto estimator of Eq.~(\ref{eq:P_yama}) is exactly captured by including the variation of the LOS in the relative weighting of different galaxies to FFTs, each covering a different axis direction, Eqs.~(\ref{eq:A2expansion}) and~(\ref{eq:A4expansion}).
 
 \section{Performance tests} \label{sec:tests}

In this section we test the following three implementations of the Yamamoto estimator, solving Eq.~(\ref{eq:P_yama}) using the following.
\begin{enumerate}
\item A sum over galaxies and randoms (the total number of points is $N$) and the $N_k$ $k$-modes of interest. We will refer to this as {\it sum-gal}.
\item A sum over a gridded representation of $F$ with $N$ grid cells, and the $N_k$ $k$-modes. We will only consider $N=N_k$ although this is not fixed as for an FFT, and refer to this as {\it sum-grid}.
\item An FFT-based implementation using a gridded representation of $F$ with $N$ grid cells and the $N_k=N$ $k$-modes. We will refer to this as {\it FFT-based}.
\end{enumerate}

For the methods using sums we have optimised our code, minimizing the computations performed within the inner most loops, and using the Hermitian symmetry in $k$-space to reduce the number of $k$-modes summed over. We also only compute power spectrum moments for $k\leq0.3h{\rm Mpc}^{-1}$ for these methods. Additionally, for the {\it sum-grid} method we only include filled grid cells in the sum. We therefore consider that time taken by these algorithms is indicative of that achieved by most algorithms performing the transform using a sum.

We will test these three options using the public mock galaxy catalogues matched to the CMASS galaxy sample of the Sloan Digital Sky Survey (SDSS-III; \citealt{eisenstein2011b}), BOSS Data Release 11 North Galactic Cap \citep{manera2013}.
These catalogues each contain approximately 525,000 galaxies.
We use the random catalogue provided with the galaxy mocks and we take the number density of the randoms to be 10 times higher than the number density of the galaxies, i.e., $\alpha^{-1}\simeq10$.
For the implementations that use a grid, we place the galaxies and randoms in a cubic box of size $L_b=3500\,{\rm Mpc}{h}^{-1}$ using the Cloud-in-Cell (CiC) prescription, to obtain the quantity $F({\bf r})$ of Eq.~(\ref{eq:FKP_factor}).
In order to correct for the effects of the grid left by the CiC scheme we have corrected appropriately by the deconvolution window proposed in \cite{jing2005}.

\begin{figure}
\centering
\includegraphics[scale=0.31]{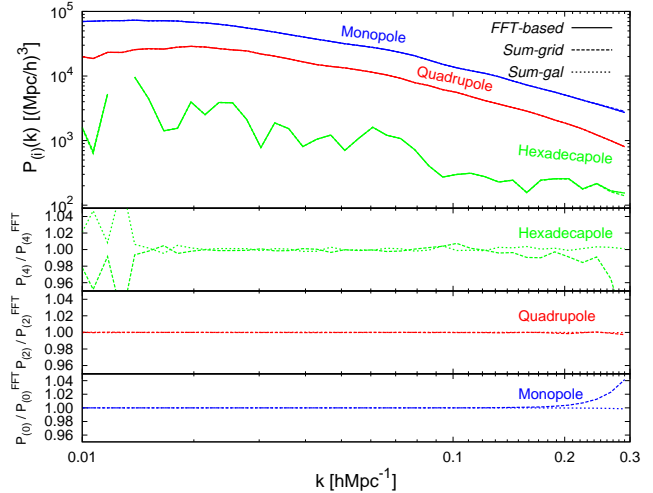}
\caption{Top panel: power spectrum multipoles: monopole (blue lines), quadrupole (red lines) and hexadecapole (green lines), obtained from the average of 50 realization of \textsc{pthalos} mocks corresponding to the BOSS DR11 CMASS NGC survey geometry. The solid lines display the computation of Eq.~(\ref{eq:P_yama}) using the {\it FFT-based} method using $1024^3$ grid cells. The dashed and dotted lines display the computation of the Yamamoto estimator using the {\it sum-grid} (with $512^3$ cells) and {\it sum-gal} methods, respectively. In both these cases an orthonormal base of $512^3$ $k$-vectors has been used. The bottom panels show the corresponding {\it sum-gal} and {\it sum-grid} multipoles divided by the {\it FFT-based} multipoles to highlight differences among these implementations.}  
\label{plot1}
\end{figure}
Fig.~\ref{plot1} displays the average power spectrum multipoles: monopole (red), quadrupole (blue) and hexadecapole (green) calculated from 50 mocks. The solid lines represent the {\it FFT-based} method, the dashed lines the {\it sum-grid}, and the dotted lines the {\it sum-gal}. The plot shows an almost exact agreement between the three implementations of Eq.~(\ref{eq:P_yama}). The results of the {\it sum-grid} algorithm show a few percent deviation at small scales. The origin of this is aliasing, which we have not corrected for. The aliasing effect for a $1024^3$ grid is negligible for scales $k\leq0.4h{\rm Mpc}^{-1}$, and consequently does not appear for the {\it FFT-based} scheme. 
For comparison, adopting a $2048^3$ grid we expect the aliasing to be negligible for wave numbers up to $\sim 0.6h{\rm Mpc}^{-1}$.
Due to its small amplitude, at small $k$ the hexadecapole is affected by numerical noise, which results in a general instability of the ratio between different methods.  
  \begin{table*}
\begin{center}
\begin{tabular}{|c|c|c|c|c|c}
 & {\it FFT-based} ($512^3$) & {\it FFT-based} ($1024^3$) & {\it FFT-based} ($2048^3$) & {\it sum-gal} ($512^3$) & {\it sum-grid} ($512^3$)\\
 \hline
 Time (min) & 1.2 & 7.5 & 72.5 & $\sim1800$ & $\sim2400$ \\
\end{tabular}
\end{center}
\caption{Computation times (in minutes, using 16 processors) for the power spectrum monopole, quadrupole and hexadecapole for the three different implementations of the Yamamoto algorithm. For the {\it FFT-based} implementation we show the number of grid cells used: $512^3$, $1024^3$ and $2048^3$. For the {\it sum-gal} algorithm the computation times are assuming $\alpha^{-1}\sim10$ and for both {\it sum-gal} and {\it sum-grid} algorithms only computing for $k\leq0.3h{\rm Mpc}^{-1}$.}
\label{table}
\end{table*}
 
In Table~\ref{table} we show a comparison between the computation times of the different algorithms of Fig.~\ref{plot1} for the monopole, quadrupole and hexadecapole of one realization of the DR11 CMASS NGC mocks.
For the {\it FFT-based} implementation, we also show the computation times for different number of cells used.
If we relax our assumption of 10 times randoms, and use $X_{\rm ran}$ times as many randoms as galaxies (for example, \citealt{anderson2014} used $X_{\rm ran} = 50$), then the computational time taken for {\it sum-gal} scales by approximately $(X_{\rm ran}+1)/11$.
For multiple measurements for different catalogues that use the same randoms, then the time in the table reduces by a factor $1/11$ for each catalogue where the randoms do not have to be reused.
However, note that in the post-reconstruction analyses of \citet{anderson2014}, the randoms are uniquely matched to each galaxy catalogue and so have to be calculated for each mock.
The {\it sum-grid} method does not scale with the number of randoms, and is therefore faster than {\it sum-gal} when the number of randoms to be analysed is larger.
Finally, when comparing run times, note that for {\it sum-gal} there is no aliasing as the galaxies and randoms are not placed on a grid, so we can use the same $N_k$ to push to smaller $k$ than the grid-based routines.
Even allowing for these scalings, it is clear that the {\it FFT-based} method is significantly faster (approximately 1000 times) than either {\it sum-gal} or {\it sum-grid} for reasonable assumptions of grid size and number of randoms. 

\section{General moments of the Power Spectrum}  \label{sec:moments}

\begin{figure}
\centering
\includegraphics[scale=0.31]{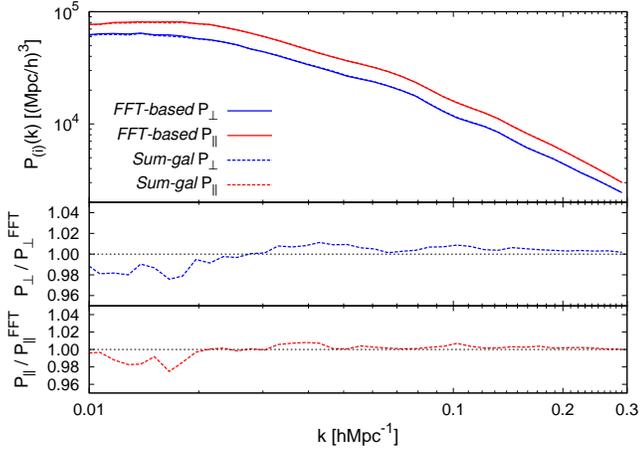}
\caption{Top panel: power spectrum `Wedges': perpendicular-to-the-LOS power spectrum monopole, $P_\perp$ (blue lines) and parallel-to-the-LOS power spectrum monopole (red lines) obtained from the average of 50 realization of \textsc{pthalos} mocks corresponding to the BOSS DR11 CMASS NGC survey geometry. The solid lines display the approximation presented by Eq.~(\ref{P_perp}-\ref{P_para}) using the monopole, quadrupole and hexadecapole computed by using the {\it FFT-based} method described in \S\ref{sec:algorithm} placing the particles in $1024^3$ grid cells. The dashed lines display the computation of the ``Wedges" using {\it sum-gal} and  Eq.~(\ref{P_perp_def}-\ref{P_para_def}), so the sum is exact. In this case an orthonormal base of $512^3$ $k$-vector have been used. The bottom panels show the fractional differences between the {\it sum-gal} and the {\it FFT-based} method, for $P_\perp$ and $P_\parallel$ as labeled.}
\label{plot2}
\end{figure}

The trick of splitting $\mu^n$ into Cartesian components employed in Eq.~(\ref{eq:kdotr}) will not work directly on moments of more general functions of $\mu$. However we can still use a {\it FFT-based} method by decomposing the functions into Legendre polynomials and summing over the multipole-moments. For example, one proposed alternative to using multipoles is to use ``Wedges'' \citep{kazin2012}, where we replace $\mathcal{L}_\ell(\mu)$ in Eq.~(\ref{eq:P_yama}) by top hat functions in $\mu$ covering $0\leq\mu\leq0.5$, whose moment we denote $P_\perp$ and $0.5<\mu\leq1$ whose moment we denote $P_\parallel$:
\begin{eqnarray}
\label{P_perp_def}P_\perp(k)&=& \frac{2}{I} \int_0^{2\pi}\frac{d\varphi}{2\pi}\int_0^{0.5}d\mu\, \left[ A_0({\bf k}) A^*_0({\bf k})\right]-P_0^{\rm noise},  \\
\label{P_para_def}P_\parallel(k)&=& \frac{2}{I} \int_0^{2\pi}\frac{d\varphi}{2\pi}\int_{0.5}^1d\mu\, \left[ A_0({\bf k}) A^*_0({\bf k})\right]-P_0^{\rm noise},
\end{eqnarray}
where $\varphi$ is the azimuthal angle.
Then, as discussed in \citet{kazin2012}, we can approximate these functions using the first three even Legendre polynomials as, 
\begin{eqnarray}
\label{P_perp}P_\perp(k)&\simeq&P_0(k)-\frac{3}{8}P_2(k)+\frac{15}{128}P_4(k), \\
\label{P_para}P_\parallel(k)&\simeq&P_0(k)+\frac{3}{8}P_2(k)-\frac{15}{128}P_4(k).
\end{eqnarray}
In Fig.~\ref{plot2} we show the comparison between the $P_\perp$ (blue lines) and $P_\parallel$ (red lines) computed using the {\it sum-gal} algorithm (dashed lines), i.e. the definition of Eq.~(\ref{P_perp_def}-\ref{P_para_def}), and the combination of Eq.~(\ref{P_perp}-\ref{P_para}) computed using the {\it FFT-based} algorithm (solid lines).  The agreement between the definition of $P_\perp$ and $P_\parallel$ and the approximation of Eq.~(\ref{P_perp}-\ref{P_para}) is very good for the range of scales studied. This suggest that the Yamamoto implementation based on FFTs presented in this Letter is also suitable to be used to compute the wedges power spectral moments.

\section{Conclusions}

We have explored methods for implementing the calculation of LOS-dependent moments of the galaxy power spectrum. Following on from developments in \citet{yamamoto2006} and \citet{blake2011} we have shown that the resulting equation can be solved using multiple FFTs, thus providing a fast method to measure LOS-dependent clustering. We have shown that this is faster than previous methods using sums over galaxies, and this will also be faster than pair-counting algorithms based on the \citet{landy1993} algorithms to calculate configuration-space monopole, quadrupole and hexadecapole moments of the correlation function. Developments such as this are necessary given the next generation of galaxy redshift surveys, including DESI \citep{levi2013} and {\it Euclid} \citep{laureijs2011}, will provide an order of magnitude more galaxies than current surveys, and therefore make computations more challenging. Developments such as that presented here should also find application in the measurement of the bispectrum, and contribute to our ability to fully exploit galaxy surveys to provide cosmological information.

After submission of our Letter and publication on the archive, a similar derivation appeared  \citep{scoccimarro2015}. This additionally showed that the hexadecapole can be calculated from the FFTs used to estimate the quadrupole, using relationships of Legendre polynomials and a slightly different LOS approximation.

\section*{Acknowledgements}
DB, RR and WJP acknowledge support from the European Research Council through grant {\it Darksurvey}. HGM and WJP acknowledge support from the UK Science \& Technology Facilities Council through the consolidated grant ST/K0090X/1, and WJP also acknowledges support from the UK Space Agency through grant ST/K00283X/1. Numerical computations were performed using the \textsc{Sciama} High Performance Compute (HPC) cluster which is supported by the ICG, SEPNet and the University of Portsmouth.\\
 
\bibliographystyle{./mn2e}
\bibliography{./biblio_db}

 \end{document}